\documentclass[english,11pt]{article}
\input epsf

\title{Counting Photons in the $\boldmath\Lambda$-Experiment}

\author{Anita D\c{a}browska\\ \\
Chair of Theoretical Foundations of Biomedical Sciences and
Medical Informatics\\
Ludwik Rydygier Medical University \\ ul.~Jagiello\'nska 13,
85-067 Bydgoszcz, Poland }

\begin{document}
\maketitle
\begin{abstract}
Dehmelt's $\Lambda$-experiment for a three-level atom with
simultaneously driven strong and weak transition is studied within
quantum stochastic calculus approach. The statistics of the
emitted photons is found by the method of generating functional of
the corresponding two dimensional output counting process. In
particular, the average waiting times for a count are calculated.
\end{abstract}

\section{Introduction}
In the eighties of last century the series of Dehmelt's papers
\cite{Deh82,Deh75} became the source of inspiration for many
authors who were fascinated by the possibility of observing the
fluorescence light emitted by a single confined atom or ion.
Dehmelt suggested a very sensitive scheme for detecting very weak
transitions in a single trapped ions (or atoms) by exploiting
electron shelving effect. The electron shelving effect appears in
atom with two transitions: one strong and one weak simultaneously
driven. When the electron is shelved on the metastable level, the
fluorescence light (resulting due to the intense transition) is
switched off. Consequently, the fluorescence light emitted by a
single atom exhibits periods of darkness. The length of the dark
periods for a three-level system in the $V$-configuration was
obtained by Cohen-Tannoudji and Dalibard~\cite{Coh86}. These
results were confirmed by Barchielli \cite{Bar87} who gave also
the physical explanation of these phenomena in the language of
quantum stochastic calculus (QSC) \cite{HuPa84,GarCo85}. Following
\cite{Bar87} we apply the mathematical theory of QSC to describe
the Dehmelt experiment for a three-level atom in the
$\Lambda$-configuration. To find the statistics of the emitted
photons we use the theory of the counting processes
[7-11]. 
The generating
functional approach of \cite{StaszewskiStaszewska93} allows us
calculate the average waiting times for the counts. Belavkin's
filtering equation for a quantum system under the counting
observation, cf.~\cite{BelL89} and the literature therein, enables
us to study the problem in terms of pure posterior quantum states.

\section{The model of experiment within QSC}
Let us consider a three-level atom with two transitions: very
intense $|1 \rangle \leftrightarrow |0\rangle$ and very weak one
$|0\rangle \leftrightarrow |2\rangle$. We will call these
transitions for convenience the ``blue'' and the ``red'' and
assume that they are driven by two lasers.
\begin{center}
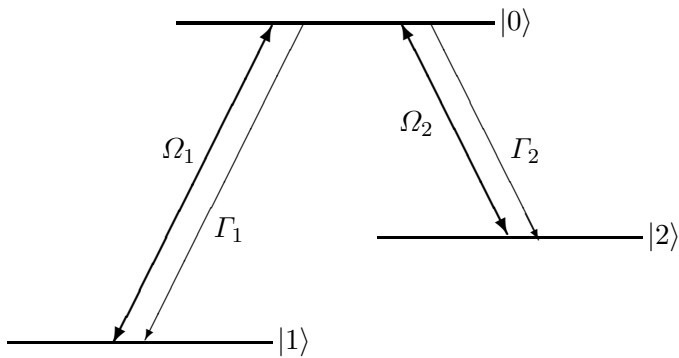
\begin{figure}[t]
\begin{picture}(500,150)(50,1)
\put(268,100){${\mit\Omega}_{2}$} \put(310,90){${\mit\Gamma}_{2}$}
\put(198,60){${\mit\Gamma}_{1}$} \put(178,90){${\mit\Omega}_{1}$}
 \put(306,138.5){$| 0 \rangle$}
\put(362,57.5){$| 2 \rangle$} \put(222,17.5){$| 1 \rangle$}
{\thicklines \put(120,20){\line(1,0){100}}}  {\thicklines
\put(184,141){\line(1,0){120}}} {\thicklines
\put(260,60){\line(1,0){100}}} {\thicklines
\put(160,20){\vector(-1,-2){0}}}
\put(280,141){\vector(1,-2){41}}
{\thicklines \put(269,141){\vector(1,-2){40}}} {\thicklines
\put(269,141){\vector(-1,2){0}}}
{\thicklines
\put(160,20){\vector(1,2){60}}} \put(232,141){\vector(-1,-2){60}}
\end{picture}
\noindent\caption{\small Energy-level scheme for the $\Lambda$
configuration.}
\end{figure}
\end{center}
\noindent The hamiltonian $H$ of the system is given by formula
\cite{Sho90}
\begin{equation}\label{dd.1}
H\;=\; \sum\limits_{k=1}^{2} \bigg[ \frac{1}{2}
{\mit\Omega}_{k}\big( | 0 \rangle \langle k | \,+\, | k \rangle
\langle 0 | \big) \,+\, {\mit\Delta}_{k} | k \rangle \langle k |
\bigg]\,,
\end{equation}
where ${\mit\Omega}_{k}$-s represent  Rabi frequencies,
${\mit\Delta}_{k}$-s are detuning parameters. According to
\cite{Bar87}, cf.~also \cite{GarCo85}, the dynamics of the system
`atom plus electromagnetic field' can be described by quantum
stochastic differential
equation (QSDE) if we recall to the following physical
approximations:
\begin{itemize}\addtolength{\itemsep}{-5pt}
\item[{\rm (i)}] the interaction atom-field is linear in the
field operators,
\item[{\rm (ii)}] the rotating-wave approximation (RWA) is
made,
\item[{\rm (iii)}] the field spectrum is flat, and the `coupling
constants' do not depend on frequency.
\end{itemize}

\noindent Approximation (i). The interaction between an atom and
electromagnetic field, in the interaction picture with
respect to the free dynamics of the field and in the dipole
approximation, takes the form
$$ -e \mbox{\boldmath $r$} \cdot \mbox{\boldmath
$E$}(\mbox{\boldmath $r$}, t) \,,$$ where \mbox{\boldmath $r$} is
the position of the electron, and \mbox{\boldmath $E$} is the
quantum electric field.

\vspace{1.5ex} \noindent Approximation (ii). Let $b_{j}(\omega)$,
$b_{j}^{\dagger}(\omega)$, $j \in I$, where $I$ is a countable
set, denote modal annihilation and creation operators satisfying
canonical commutation relations
$[b_{j}(\omega),b_{i}^{\dagger}(\omega')]=\delta_{ji}
\delta(\omega-\omega')$, where $\omega\geq0$ is a continuous index
representing energy, and $j$ is a discrete index. The
positive-frequency electric field operator $\mbox{\boldmath
$E$}^{+}(\mbox{\boldmath $r$}, t)$ can be written in terms of
$b_{j}(\omega)$-s as
\begin{equation}\label{dd.2}
\mbox{\boldmath $E$}^{+}(\mbox{\boldmath $r$}, t)\;=\; \sum_{j}
\,(2\pi)^{-1/2} \int_{0}^{+\infty} \mbox{\boldmath
$F$}_{j}(\mbox{\boldmath $r$}, \omega) {\rm e}^{-i\omega t}
b_{j}(\omega)\, d\omega  \,.
\end{equation}
The explicit form of the coefficients $\mbox{\boldmath
$F$}_{j}(\mbox{\boldmath $r$},\omega)$ is not important for our
discussion, we stress only that index $j$  also contains the
direction of propagation of photons. Let $\psi_{0}(\mbox{\boldmath
$r$})$, $\psi_{k}(\mbox{\boldmath $r$})$ (k=1,\,2) denote the wave
functions of the considered states. The interaction hamiltonian in
the RWA can be written as

$$ H_{k0} = -e | k
\rangle \int \bar{\psi}_{k}(\mbox{\boldmath $r$}) \mbox{\boldmath
$r$} \cdot \mbox{\boldmath $E$}^{+}(\mbox{\boldmath $r$}, t)
\psi_{0}(\mbox{\boldmath $r$})\, d^{3} \mbox{\boldmath $r$} \,
\langle 0 | \,+\, \mbox{\rm hc}\,.$$ If we introduce

$$ f_{j}^{k0}
\;:= \; -e \int \bar{\psi}_{k}(\mbox{\boldmath $r$})
\mbox{\boldmath $r$} \cdot \mbox{\boldmath
$F$}_{j}(\mbox{\boldmath $r$}, \omega) \psi_{0}(\mbox{\boldmath
$r$})\,d^{3} \mbox{\boldmath $r$}\,,
$$ then the interaction hamiltonian takes the form

$$ H_{k0} \;=\; | k \rangle \sum_{j} \, (2 \pi)^{-1/2}
\int_{0}^{+\infty} {\rm e}^{-i\omega t} f_{j}^{k0}(\omega)
b_{j}(\omega)\, d \omega \, \langle 0 | \,+\, \mbox{\rm hc}\,.
$$
Therefore, in the interaction picture with respect to free
dynamics of the atom, one gets

$$H_{k0} \;=\; | k \rangle \sum_{j}
\, (2 \pi)^{-1/2} \int_{-\omega_{k}}^{+\infty}  {\rm e}^{-i\omega
t } f_{j}^{k0}(\omega+\omega_{k} ) b_{j}(\omega+\omega_{k}) \, d
\omega \, \langle 0 | \,+\, \mbox{\rm hc} \,,
$$ where $\omega_{k}$, $k=1,\,2$, denote the considered
frequencies.

\vspace{1.5ex}
\noindent Approximation (iii). We assume that $f^{k0}_{j}
(\omega)$ is constant in a neighborhood of $\omega_{k}$ and zero
elsewhere, therefore the expression $f^{k0}_{j} (\omega +
\omega_{k})$ can be replaced with $f^{k0}_{j} (\omega_{k})$
(coupling constant independent of frequency). The range of
integration can be extended from $- \infty$ to $+ \infty$
according to the fact that the field  spectrum is flat. Hence
$$ H_{k0} \;=\; | k \rangle \langle 0 | \sum_{j} \,
f_{j}^{k0}(\omega_{k}) b_{j}(t) \,+\, \mbox{\rm hc} \,,$$ where
\begin{equation}\label{ani}
b_{j}(t) \;=\; (2 \pi)^{-1/2} \int_{-\infty}^{+\infty}
 {\rm e}^{-i\omega t} b_{j}(\omega+ \omega_{k}) \, d\omega
\,.
\end{equation}

\noindent Let us denote the disjoint sets of indices labeling
independent field modes carrying blue and red photons,
respectively by $I_{1}$, $I_{2}$. Then the coupling systematic
operators are given by the expressions
\begin{equation}\label{dd.3}
L_{j} \;=\; \left\{ \begin{array}{ll}
z_{j} S_{1} \;=\; z_{j}\, | 1 \rangle \langle 0 | & \,\mbox{if
$\quad j \in I_{1}$\,,} \\
z_{j} S_{2} \;=\; z_{j} \, | 2 \rangle \langle 0 | & \, \mbox{if
$\quad j \in I_{2} $\,,}
\end{array}
\right.
\end{equation}
where $z_{j}$-s are complex coupling constants and $S_{j}$-s are
lowering atomic operators.

\noindent Let us introduce the stochastic differentials of the
annihilation and creation processes:
\begin{eqnarray}\label{diff}
dB(t)_{j}&=&B(t+dt)-B(t) \;=\; \int_{t}^{t+dt} b_{j}(t^{\prime}) \,
dt^{\prime}\,,\nonumber \\
dB^{\dagger}(t)_{j}&=&B^{\dagger}(t+dt)-B^{\dagger}(t) \;=\;
\int_{t}^{t+dt} b_{j}^{\dagger}(t^{\prime}) \,dt^{\prime} \,.
\end{eqnarray}
With the help of these differentials one can represent the
dynamical evolution equation for the unitary
evolution operator for `the system plus electromagnetic field' in
the form of the Ito quantum stochastic differential equation
 (QSDE) of the form \cite{HuPa84,GarCo85}
\begin{equation}\label{unitary}
dU(t)\;=\;-KU(t) \, dt+ \sum_{k=1}^{2} \sum_{j \in I_{k}}
 \left( L_{j}
dB_{j}^{\dagger}(t) - L_{j}^{\dagger}dB_{j}(t)
\right) U(t) \,, \qquad U(0) \,=\, I\,,
\end{equation}
where
\begin{equation}\label{ka}
K\;=\;iH+ \frac{1}{2} \sum_{k=1}^{2} \sum_{j \in I_{k}} \,
 L_{j}^{\dagger}L_{j}
\end{equation}
and $H$ is given by the formula (\ref{dd.1}). With the help of
(\ref{dd.3}) one obtains
\begin{equation}\label{dd.11}
K \;=\; i H + \frac{1}{2} \, ({\mit\Gamma}_{1}+{\mit\Gamma}_{2})\,
| 0 \rangle \langle 0 |\,,
\end{equation}
where
\begin{equation}\label{dd.8}
{\mit\Gamma}_{k}=\sum_{j \in I_{k}} |{z}_{j}|^{2}\,\,,\qquad
k=1,\,2
\end{equation}
and the quantities ${\mit\Gamma}_{k}$ stand for total transition
rates for the blue and red transitions.

\section{The statistics of the output counting process}

\noindent Let us consider two-dimensional photon counting process
\begin{equation}\label{dd.4}
 \widehat{{\mbox{\boldmath
$\mathcal{N}$}}}(t)=[\widehat{\mathcal{N}}_{j}(t)]_{j=1}^ {2}
\end{equation}
with components
\begin{equation}\label{dd.5}
\widehat{\mathcal{N}}_{k}(t) \;:=\; \sum\limits_{j \in I_{k}}
\int\limits_{0}^{t} \, \widehat{b}_{j}^{\dagger}(t^{\prime})
\widehat{b}_{j}(t^{\prime}) \, d t^{\prime} \,, \qquad k=1,\,2,
\end{equation}
connected with blue and red photons. Here
\begin{eqnarray}\label{beout}
\widehat{b}_{j}(t)&=& U(t)^{\dagger}b_{j}(t)U(t)\,,\nonumber \\
\widehat{b}^{\dagger}_{j}(t)&=& U(t)^{\dagger}b_{j}(t)U(t)\,
\end{eqnarray}
stand for the output field annihilation and creation operators
 \cite{GarCo85}, cf.~also \cite{Bar86}. These describe the
field after interacion with the atom within the time interval
$(0,\,t)$. Therefore the counting process is called the output
counting process. \noindent We assume that all emitted photons are
detected.

\noindent If the atom is prepared initially in a pure state, then
 the linear
Belavkin filtering equation for the
posterior unnormalized wave function takes the form \cite{BelL89}
\begin{equation}\label{dd.6}
d \widehat{\varphi} (t)\;=\; - \left( iH + \frac{1}{2}
{\mit\Gamma} | 0\rangle \langle 0 | - \frac{1}{2} {\mit\Gamma}
\right)
 \widehat{\varphi}(t) \, dt \,+\, \sum_{j=1}^{2} \, ( S_{j} - I )
 \widehat{\varphi}(t)d\widehat{\mathcal{N}}_{j}(t)\,,
\end{equation}
where
\begin{equation}\label{dd.7}
{\mit\Gamma} \;:= \; \sum_{k=1}^{2} {\mit\Gamma}_{k}\,.
\end{equation}
QSDE (\ref{dd.6}) plays the role analogous to that of the
Schr\"{o}dinger equation
for unobserved quantum system. It describes the time-evolution of
 a pure quantum state of the atom evolving according to the
trajectory of the counting process. (For the case of observation
process of the Wiener type see for instance \cite{Bel92} and the
literature quoted therein.)

In order to obtain the  average waiting times for the counts we do
not need to solve the equation (\ref{dd.6}). These can be deduced
from the statistics of the counting process. It is well-known that
the whole statistics of the counting process can be found by
solving the differential equation for the generating (or
characteristic) functional of the process. Here we use the results
of \cite{StaszewskiStaszewska93} in which the generating
functional approach was applied.

Let us consider the counting trajectory up to $t$, $\kappa = \big(
(j_{m_{1}}, t_{1}),\, (j_{m_{2}}, t_{2}),\, \ldots \,,\,
(j_{m_{n}}, t_{n}) \big)$. The probability density of two or more
counts at the same time vanishes. It follows from that the
detection of a photon projects the atom into one of the lower
states, so the atom must be re-excited before the following photon
is detected. To calculate the probability density (with respect to
$ \prod_{j=1}^{n} \, dt_{j}$) of counting a photon of type
$j_{m_{1}}$ at time $t_{1}$, a photon of type $j_{m_{2}}$ at time
$t_{2}$, \, \ldots \, a photon of type $j_{m_{n}}$ at time
$t_{n}$, where $t_{1} < t_{2} , \, \ldots \, , \, t_{n} <t$, and
no other photon in the time interval $(0,\,t]$ we use the formula
\cite{StaszewskiStaszewska93}
\begin{equation}\label{dd.9}
p(\kappa \,|\, t) \;=\; \prod_{j=1}^{n}{\mit\Gamma}_{m_{j}}
\langle {V}(\kappa \,|\, t) \psi \,|\,{V}(\kappa \, |\, t) \psi
\rangle\,,
\end{equation}
where
\begin{equation}\label{dd.10}
{V}(\kappa \,|\, t)\;=\; \mbox{\rm e}^{-Kt}
 S_{m_{n}}(t_{n}) \, \ldots \,
 S_{m_{1}}(t_{1}) \,,
\end{equation}
\begin{equation}\label{dd.12}
S_{m}(t)\;=\; \mbox{\rm e}^{Kt} S_{m} \mbox{\rm e}^{-Kt}\,,
\end{equation}
and $K$ is given by (\ref{ka}).

\noindent And hence, if we assume that atom was in the state
$|1\rangle$ at the initial moment $t=0$
\begin{equation}\label{dd.13}
p((j_{m_{1}},t_{1})\,|\,t) \;=\; P_{t_{1}}^{t}(0 \, |\,|k
\rangle)\, \Gamma_{k}| \langle 0 | e^{-Kt_{1}} | 1 \rangle
|^{2}\,,
\end{equation}
where the quantity
\begin{equation}\label{dd.14}
P_{t_{1}}^{t}(0 \,|\,|k\rangle) \;=\; \parallel \mbox{\rm
e}^{-K(t-t_{1})} \,|\, k \rangle \parallel^{2}
\end{equation}
is the probability of having no counts within the time-interval
$(t_{1}, \, t]$ for the atom being in the state $| k \rangle$
$(k=1,\,2)$ at the instant $t_{1}$.

\noindent From (\ref{dd.13}) we obtain the probability density of
one count within the time-interval $(t_{0}, \, t]$ conditioned by
having a count of a photon of the type $j_{0}$ at $t_{0}$
\begin{equation}\label{dd.15}
p_{t} \big( (j_{m_{1}}, \, t_{1}) \, | (j_{0}, \, t_{0}) \big)
\;=\; p_{t-t_{0}} \big( (j_{m_{1}} , \, t_{1}-t_{0}) \,|\, |k
\rangle \big) \qquad k = 1,\,2.
\end{equation}
The form of the above expression is compatible with fact that
after a count of a blue photon the atom is in the $|1 \rangle$
state and after a count of a red photon the atom is in the $|2
\rangle$ state.

\noindent The structure of (\ref{dd.15}) allows us write down the
probability density of detecting a blue or red photon  at the time
$t_{0} + t$ and no other photons  in the time-interval $( t_{0},
\,t_{0} + t] $, conditioned by having a count a blue photon at the
instant $t_{0}$
\begin{equation}\label{dd.16}
W_{k}(t|\, |1 \rangle) \;=\; \Gamma_{k} | \langle 0 |
e^{-K(t-t_{0})} | 1 \rangle |^{2}\,, \qquad k = 1,\,2.
\end{equation}
If we assume that a red photon was detected at the moment
$t_{0}$ then
\begin{equation}\label{dd.17}
W_{k}(t|\, |2 \rangle) \;=\; \Gamma_{k} | \langle 0 |
e^{-K(t-t_{0})} | 2 \rangle |^{2}\,, \qquad k = 1,\,2,
\end{equation}
represents the  conditional probability density of detecting a
blue or red photon at the instant $t_{0} + t$ and no other photons
in the time-interval $( t_{0}, \,t_{0} + t]$.

\section{Existence of dark periods}

To describe phenomena occuring in the $\Lambda$-experiment we will
follow the approach of \cite{Coh86} and \cite{Bar87}. The authors
of \cite{Coh86} noticed that the characteristics of dark periods
is controlled by the probabilities $P_{t_{0}}^{t_{0}+t} (0 \, |\,
|k \rangle)$ for not having any count in the time-interval
$(t_{0},t_{0}+t]$, proceeded by the count of photon at instant
$t_{0}$. Due to (\ref{dd.15}) one can put for convenience
$t_{0}=0$. We will also simplify the notation by writing $P (t \,
|\, |k \rangle)$ instead of $P_{0}^{t} (0 \, |\, |k \rangle)$. To
find these probabilities one should calculate probability
densities $W_{j}(t|\, |k \rangle)$. For this purpose let us set
\begin{equation}\label{dd.18}
\mbox{\rm e}^{- K t} | k \rangle \;=\; \sum\limits_{j=0}^{2}
a_{j}(t|\, |k \rangle) | j \rangle \,, \;\;k = 1,\,2.
\end{equation}

\noindent By formulas (\ref{dd.16}) and (\ref{dd.17}) one gets
\begin{equation}\label{dd.19}
W_{j}(t|\, |k \rangle) \;=\; {\mit\Gamma}_{j} |a_{0}(t|\, |k
\rangle)|^2\, \;\;k =1,2\;\;\;j =1,\,2\,.
\end{equation}

\noindent The coefficients $a_{j}(t| \, |k \rangle)$ satisfy the
equations of motion:
\begin{eqnarray}
\dot{a}_{1}(t| \,|k \rangle) &=& - \frac{i}{2} {\mit\Omega}_{1}
a_{0}(t|\, |k \rangle) \,-\, i {\mit\Delta}_{1} a_{1}(t|\, |k
\rangle)\,, \nonumber
\\ \dot{a}_{0}(t|\, |k \rangle) &=&  - \frac{i}{2}
{\mit\Omega}_{1} a_{1}(t|\, |k \rangle) \,-\,
\frac{i}{2}{\mit\Omega}_{2} a_{2}(t|\, |k
\rangle)\,-\,\frac{1}{2}{\mit\Gamma}
a_{0}(t|\, |k \rangle)\,, \label{dd.20} \\
\dot{a}_{2}(t|\, |k \rangle) &=& - \frac{i}{2} {\mit\Omega}_{2}
a_{0}(t|\, |k \rangle) \,-\, i {\mit\Delta}_{2} a_{2}(t|\, |k
\rangle) \nonumber
\end{eqnarray}
with the initial conditions:
\begin{equation}\label{21}
a_{j}(0| \,|k \rangle) \;=\; \delta_{jk}\,, \qquad k=1,\,2\,\qquad
j=0,\,1,\,2.
\end{equation}

\noindent The solution to the system (\ref{dd.20}) can be obtained
by using the Laplace transform technique. To develop formulas
(\ref{dd.19}) one needs only a knowledge of
\begin{eqnarray*}
a_{0}(t| \,|1 \rangle)\!\!\!\! &=&\!\!\!\! - \frac{i}{2}
{\mit\Omega}_{1} \! \bigg( \! \frac{z_{1}\! + \! i
{\mit\Delta}_{2}} {(z_{2}\!-\!z_{1})(z_{3}\!-\!z_{1})}\, {\rm
e}^{z_{1}t} \!+\! \frac{z_{2}\! +\! i
{\mit\Delta}_{2}}{(z_{1}\!-\!z_{2})(z_{3}\!-\!z_{2})}\, {\rm
e}^{z_{2}t} + \frac{z_{3}\! +\! i
{\mit\Delta}_{2}}{(z_{1}\!-\!z_{3})(z_{2}\!-\!z_{3})}\,
{\rm e}^{z_{3}t}\! \bigg)\,,\\
a_{0}(t| \,|2 \rangle)\!\!\!\! &=&\!\!\!\! - \frac{i}{2}
{\mit\Omega}_{2} \! \bigg( \! \frac{z_{1}\! + \! i
{\mit\Delta}_{1}}{(z_{2}\!-\!z_{1})(z_{3}\!-\!z_{1})}\, {\rm
e}^{z_{1}t} \!+\! \frac{z_{2}\! +\! i
{\mit\Delta}_{1}}{(z_{1}\!-\!z_{2})(z_{3}\!-\!z_{2})}\, {\rm
e}^{z_{2}t} + \frac{z_{3}\! +\! i
\Delta_{1}}{(z_{1}\!-\!z_{3})(z_{2}\!-\!z_{3})}\, {\rm
e}^{z_{3}t}\! \bigg)\,,
\end{eqnarray*}
where $z_{j}$ $(j=1,\,2,\,3)$ are the roots of the characteristic
equation of the system (\ref{dd.20}). The others coefficients have
an analogous form.

\noindent If we assume that
\begin{equation}\label{dd.22}
{\mit\Omega}_{1} \gg {\mit\Omega}_{2}\,, \qquad  {\mit\Gamma}_{1}
\gg {\mit\Gamma}_{2}\,, \qquad {\mit\Gamma}_{1} \gg
{\mit\Omega}_{2}\,,
\end{equation}
then the roots $z_{j}$ are approximately given by
\begin{eqnarray}
z_{1} & = &  - i {\mit\Delta}_{2}+\zeta\,, \label{dd.23}
\\z_{2} & \simeq & - \frac{1}{4}
{\mit\Gamma}-\frac{1}{2}i{\mit\Delta}_{1} + \frac{1}{4}
({\mit\Gamma}^{2} - 4 {\mit\Omega}_{1}^{2}-4{\mit\Delta}_{1}^{2}
-4i{\mit\Gamma}{\mit\Delta}_{1})^{1/2} \,, \label{dd.24}
\\z_{3} & \simeq & - \frac{1}{4}
{\mit\Gamma}-\frac{1}{2}i{\mit\Delta}_{1} - \frac{1}{4}
({\mit\Gamma}^{2} - 4 {\mit\Omega}_{1}^{2}-4{\mit\Delta}_{1}^{2}
-4i{\mit\Gamma}{\mit\Delta}_{1})^{1/2} \,,\label{dd.25}\\
\zeta & \simeq &
{\mit\Omega}_{2}^{2}({\mit\Delta}_{2}-{\mit\Delta}_{1})
\frac{i({\mit\Omega}_{1}^{2} - 4
{\mit\Delta}_{2}^{2}+4{\mit\Delta}_{1}{\mit\Delta}_{2}) -
2{\mit\Gamma}_{1}({\mit\Delta}_{2}- {\mit\Delta}_{1})
}{({\mit\Omega}_{1}^{2}-4
{\mit\Delta}_{2}^{2}+4{\mit\Delta}_{1}{\mit\Delta}_{2})^{2}+
4{\mit\Gamma}_{1}^{2}({\mit\Delta}_{2}-{\mit\Delta}_{1})^{2}}
.\,\label{dd.zeta}
\end{eqnarray}

\noindent One can observe that the real parts of the roots $z_{j}$
are negative for ${\mit\Delta}_{1} \neq {\mit\Delta}_{2}$. It can
be proved that for ${\mit\Delta}_{1} \neq {\mit\Delta}_{2}$
\begin{equation}\label{dd.26}
\sum_{j=1}^{2}\, \int_{0}^{+\infty} \,  W_{j}(t|\, |k \rangle) \,
dt \;=\; 1\,\qquad k=1,\,2\,.
\end{equation}

\noindent This means, that the probability that at least one
photon is detected in the interval $(0,+\infty)$ is one if
${\mit\Delta}_{1} \neq {\mit\Delta}_{2}$ (we have assumed that all
emitted photons are detected). To prove of (\ref{dd.26}) let us
notice that
\begin{equation}
\sum_{j=1}^{2} \, W_{j}(t| |k \rangle) \;=\; \mbox{\rm
Tr}\kern.7pt \big[ (K+K^{\dagger}) \mbox{\rm e}^{-Kt} |k\rangle
\langle k| \mbox{\rm e}^{-K^{\dagger}t} \big] \;=\; - \frac{d}{dt}
\mbox{\rm Tr}\kern.7pt \big[ \mbox{\rm e}^{-Kt} |k\rangle \langle
k| \mbox{\rm e}^{-K^{\dagger}t} \big] \,,\label{dd.27}
\end{equation}
\begin{equation}
\sum_{j=1}^{2} \int_{0}^{+\infty} \, W_{j}(t| |k \rangle) \,dt
\;=\; 1 \,-\, \lim_{t \rightarrow +\infty}  \mbox{\rm Tr}\kern.7pt
\big[ \mbox{\rm e}^{-Kt} |k\rangle \langle k| \mbox{\rm
e}^{-K^{\dagger}t} \big]\,. \label{dd.28}
\end{equation}

\noindent The limit appearing on the right-hand side of
(\ref{dd.28}) vanishes provided the real parts of the roots
$z_{j}$ are negative. Thus, (\ref{dd.26}) is satisfied if
${\mit\Delta}_{1} \neq {\mit\Delta}_{2}$. \\

\noindent Let us now consider the case of
\begin{equation}
{\mit\Delta}_{1} = {\mit\Delta}_{2} = {\mit\Delta}.
\,\label{dd.29}
\end{equation}

\noindent The condition (\ref{dd.29}) implies that the difference
in photon energies matches the energy separation between the two
lower atomic states.  One can say that this is the condition for a
Raman-type two-photon transition between states $|1 \rangle$ and
$|2 \rangle$. The roots $z_{j}$ of characteristic equation of the
system (\ref{dd.20}) have now the form
\begin{eqnarray}
z_{1} & = &  -i{\mit\Delta}, \label{dd.30}
\\z_{2} & = & - \frac{1}{4} {\mit\Gamma}-\frac{1}{2}i{\mit\Delta}+
\frac{1}{4} ( {\mit\Gamma}^{2} - 4 {\mit\Omega}_{1}^{2} - 4
{\mit\Omega}_{2}^{2}-4{\mit\Delta}^2-4i{\mit\Gamma}{\mit\Delta})^{1/2}
\,, \label{dd.31}
\\z_{3} & = & - \frac{1}{4} {\mit\Gamma}-\frac{1}{2}i{\mit\Delta} -
\frac{1}{4} ( {\mit\Gamma}^{2} - 4 {\mit\Omega}_{1}^{2} - 4
{\mit\Omega}_{2}^{2}-4{\mit\Delta}^2-4i{\mit\Gamma}{\mit\Delta})^{1/2}
\,. \label{dd.32}
\end{eqnarray}

\noindent Therefore, one can get
\begin{equation}\label{dd.33}
\sum_{j=1}^{2}\, \int_{0}^{+\infty} \,  W_{j}(t|\, |1 \rangle) \,
dt \;=\; \frac{{\mit\Omega}_{1}^{2}}{{\mit\Omega}_{1}^{2} +
{\mit\Omega}_{2}^{2}}\,,
\end{equation}
and
\begin{equation}\label{dd.34}
\sum_{j=1}^{2}\, \int_{0}^{+\infty} \,  W_{j}(t|\, |2 \rangle) \,
dt \;=\; \frac{{\mit\Omega}_{2}^{2}}{{\mit\Omega}_{1}^{2} +
{\mit\Omega}_{2}^{2}}\,.
\end{equation}

\noindent To calculate (\ref{dd.33}) and (\ref{dd.34}) we have not
taken into account the assumption (\ref{dd.22}). For
${\mit\Omega}_{1} \gg {\mit\Omega}_{2}$,  the probability that
after emission of a red photon at $t_{0}=0$ at least one photon is
emitted in the interval $(0,+\infty)$, is much less than one. This
means that in this case dark periods are extremely long or
infinite. Moreover, it is worth to notice that from (\ref{dd.33})
and (\ref{dd.34}) it follows that there exist the possibility that
the atom does not ever emit any photon. In other words, the
population can be forever trapped.

\noindent Let us now analyse the situation when the condition
(\ref{dd.29}) for equal detunings is not satisfied and assume that
a blue photon was counted at the instant $t_{0}=0$. According to
(\ref{dd.11}), (\ref{dd.14}), (\ref{dd.16}) one has
\begin{eqnarray}
&& P(0|\,|1 \rangle) \;=\; 1\,,  \label{dd.35}
\\ && \frac{d}{dt} \, P(t|\, |1 \rangle) \;=\; - \sum_{j=1}^{2}
\, W_{j}(t|\, |1 \rangle) \,. \label{dd.36}
\end{eqnarray}
Therefore, one gets
\begin{equation}\label{dd.37}
P(t|\,|1 \rangle)\,=\, 1 \,-\, \sum_{j=1}^{2} \, \int_{0}^{t} \,
W_{j}(t^{\prime}|\, |1 \rangle) \, dt^{\prime}\,.
\end{equation}

\noindent Due to (\ref{dd.26}) one can write
\begin{equation}
P(t|\, |1 \rangle)\,=\, \sum_{j=1}^{2} \, \int_{t}^{+\infty} \,
W_{j}(t^{\prime}|\, |1 \rangle) \, dt^{\prime}\,. \label{dd.38}
\end{equation}

\noindent Assuming that a red photon was counted at the instant
$t_{0}=0$, one gets

\begin{equation}
P(t|\, |2 \rangle)\,=\, \sum_{j=1}^{2} \, \int_{t}^{+\infty} \,
W_{j}(t^{\prime}|\, |2 \rangle) \, dt^{\prime}\,. \label{dd.39}
\end{equation}

\noindent Since the expression $1-P(t| | k\rangle)$ is the
probability of at least one count in $(0,t]$ the quantities
\begin{equation}\label{dd.40}
p(t|\, |k \rangle)\;:=\;\frac{d}{dt}[1-P(t|\, |k \rangle)
]\;=\;F_{\scriptsize\rm short}(t|\, |k
\rangle)+F_{\mbox{\scriptsize\rm long}}(t|\, |k \rangle)
\end{equation}
are probability densities for a waiting time for a count after
emission respectively of a blue photon and a red one.

\noindent Let us consider the case when a  blue photon was emitted
at the moment $t_{0}=0$. One can write

\begin{equation}\label{41}
F_{\mbox{\scriptsize\rm long}}(t|\,|1 \rangle)\;
=\;\frac{{\mit\Omega}_{1}^2{\mit\Gamma}|z_{1}+i{\Delta}_{2}|^{2}\exp(2{\rm
Re}(z_1)t)}{4|(z_{2}-z_{1})(z_{3}-z_{1})|^{2}}.
\end{equation}

\noindent The roots $z_{j}$ are now given by the expressions
(\ref{dd.23})--(\ref{dd.zeta}). We assume additionally that
${\mit\Delta}_{1}=0$. From (\ref{dd.22}) it follows that
\begin{equation}\label{dd.42}
 |\mbox{\rm Re}\kern.7pt
z_{1} | \;\ll \; |\mbox{\rm Re}\kern.7pt z_{2,\,3} |\,.
\end{equation}

\noindent Let us set
\begin{equation}\label{dd.43}
{\mit\Pi} \;:=\;\int^{+\infty}_0 \, F_{\mbox{\scriptsize\rm
long}}(t|\,|1 \rangle) \, dt\,,
\end{equation}
then by (\ref{dd.38}) and (\ref{dd.40}) we obtain
\begin{equation}\label{dd.44}
\int_0^{+\infty}\, F_{\scriptsize\rm short}(t|\,|1 \rangle)\, dt
\;=\;1-{\mit\Pi} \,.
\end{equation}
With the help of (\ref{dd.23}) -- (\ref{dd.zeta}) one gets
\begin{eqnarray}
F_{\scriptsize\rm short}(t|\,|1 \rangle) &\simeq&
\frac{{\mit\Omega}^2_1{\mit\Gamma}}{|{\mit\Gamma}^2
-4{\mit\Omega}_1^2|}|e^{z_2t}-e^{z_3t}|^2 \,, \label{dd.45}
\\ F_{\mbox{\scriptsize\rm long}}(t|\,|1 \rangle)&=&{\mit\Pi} |2 {\rm Re} (z_1)|\exp[2{\rm
Re}(z_1)t]\,\label{dd.46}
\\{\mit\Pi}  & \simeq & \, \frac{{\mit\Omega}_{1}^{2}{\mit\Omega}_{2}^{2}}
{({\mit\Omega}_{1}^{2} - 4{\mit\Delta}_{2}^{2})^{2} + 4
{\mit\Gamma}^{2}{\mit\Delta}_{2}^{2}}.\label{dd.47}
\end{eqnarray}

\noindent Thus, from (\ref{dd.22}) it follows that ${\mit\Pi} \ll
1$. We introduce a time delay $\theta$ satisfying the following
relation
\begin{equation}\label{dd.48}
|2{\rm Re}(z_2,_3)|^{-1}\ll\theta\ll |2{\rm Re}(z_1)|^{-1}\,.
\end{equation}

\noindent The time-interval ${\mit\Delta}t$ between two successive
counts is considered as short, if ${\mit\Delta}t < \theta$, and as
long, if ${\mit\Delta}t > \theta$. The probability of a short
waiting time after the emission of a blue photon is
\begin{equation}\label{dd.49}
P({\mit\Delta} t <\theta\|1 \rangle)\;=\;\int_0^\theta \, p(t|\,|1
\rangle) \, dt \; \simeq \; \int_0^\theta \, F_{\scriptsize\rm
short}(t|\,|1 \rangle) \, dt \; \simeq \; \int_0^{+\infty}\,
F_{\scriptsize\rm short}(t|\,|1 \rangle) \, dt \;=\; 1 - {\mit\Pi}
\,,
\end{equation}
and the probability of a long waiting time reads
\begin{equation}\label{dd.50}
P({\mit\Delta} t>\theta\|1 \rangle) \;\simeq \; \int_0^{+\infty}
\, F_{\mbox{\scriptsize\rm long}}(t|\,|1 \rangle) \, dt
\;=\;{\mit\Pi}\,.
\end{equation}
According to (\ref{dd.22}), none of these probabilities depend on
$\theta$. The short waiting times are distributed with a
probability density $(1- {\mit\Pi})^{-1} F_{\mbox{\scriptsize\rm
short}}(t|\,|1 \rangle)$ and the long ones with a probability
density ${\mit\Pi}^{-1}F_{\mbox{\scriptsize\rm long}}(t|\,|1
\rangle)$. Hence, the mean duration $T_{\mbox{\scriptsize\rm
short}}$ of the short waiting intervals after the emission of a
blue photon has the form
\begin{equation}\label{dd.51}
T_{\mbox{\scriptsize\rm short}} \;=\; \frac{1}{1-{\mit\Pi}} \,
\int_{0}^{+\infty} \, t \, F_{\mbox{\scriptsize\rm short}}(t|\,|1
\rangle) \, dt \; \simeq \; {\mit\Gamma}{\mit\Omega}_{1}^{-2}
\,+\, 2 {\mit\Gamma}^{-1} \,.
\end{equation}
\noindent The mean duration $T_{\mbox{\scriptsize\rm long}}$ of
the long waiting intervals after the emission of a blue photon is
given by the formula
\begin{equation}\label{dd.52}
T_{\mbox{\scriptsize\rm long}} \;=\; \frac{1}{{\mit\Pi}} \,
\int_{0}^{+ \infty} \, t \, F_{\mbox{\scriptsize\rm long}}(t|\,|1
\rangle) \, dt \;=\; |2 \mbox{\rm Re}\kern.7pt z_{1 } |^{-1} \,.
\end{equation}
\noindent Using the assumption (\ref{dd.22}), one obtains in a
similar way
\begin{eqnarray}
F_{\scriptsize\rm short}(t|\,|2 \rangle) &\simeq&
\frac{{\mit\Omega}^2_2{\mit\Gamma}}{4} \bigg|
\frac{z_{2}}{(z_{1}-z_{2})(z_{3}-z_{2})}e^{z_2t}+\frac{z_{3}}{(z_{1}-z_{3})(z_{2}-z_{3})}e^{z_3t}
\bigg|^2 \,, \label{dd.53}
\\ F_{\mbox{\scriptsize\rm long}}(t|\,|2 \rangle)&=&\frac{{\mit\Omega}_{2}^{2}{\mit\Gamma}|z_{1}|^2}
{4|(z_{2}-z_{1})(z_{3}-z_{1})|^2} \exp[2{\rm
Re}(z_1)t]\,.\label{dd.54}
\end{eqnarray}
\noindent It follows from (\ref{dd.22}) that
$F_{\mbox{\scriptsize\rm short}}(t|\,|2 \rangle)$ is negligible.
This proves that short waiting times after a count of a red photon
are extremely infrequent. Simple calculation yields an analogous
expression for the mean duration of the long intervals as in the
case of a blue photon detected at the initial instant $t_{0}=0$.
The difference lies in the fact that in this case long
time-intervals play the main role in describing the time evolution
of the atom.

\section{Conclusion}
We have shown that there exist periods of darkness in the
fluorescence light emitted by a three-level atom in the $\Lambda$
configuration. But in contrast to the $V$-case, the atom in the
$\Lambda$-configuration can stay in the coherent states for
extremely long or infinite time. This statement is consistent with
the corresponding results of \cite{Sho90} and  \cite{PLK86}. The
dynamics of population trapping in $\Lambda$-system has a long
history. Let us stress that our main purpose was not to give new
physical results, but to show how the theory of the counting
processes can be used to find the statistical properties of light
emitted in the $\Lambda$-experiment.

\section{Acknowledgment}

I thank Professor P. Staszewski for critical reading of the paper.

\end{document}